\documentclass[12pt]{iopart}

\usepackage[colorlinks,pdfusetitle,urlcolor=blue,citecolor=blue,linkcolor=blue,bookmarksnumbered,plainpages=false]{hyperref}

\usepackage{color}
\usepackage{soul}
\usepackage{amssymb}
\usepackage{amsfonts}

\usepackage{bm}
\usepackage{nicefrac}
\usepackage{siunitx}

\usepackage{graphicx}
\usepackage{bm}
\usepackage{braket}
\usepackage{mathrsfs}
\usepackage[normalem]{ulem}

\newcommand{\B}[1]{\mathbf{#1}}
\newcommand{\eqref}[1]{(\ref{#1})}

\begin{document}

\title{A generalized bag-like boundary condition for fields with arbitrary spin}

\author{Adam Stokes and Robert Bennett}

\address{Department of Physics \& Astronomy, University of Leeds, LS2 9JT, United Kingdom}

\date{\today}

\begin{abstract} Boundary conditions for the Maxwell and Dirac fields at material surfaces are widely-used and physically well-motivated, but do not appear to have been generalised to deal with higher spin fields. As a result there is no clear prescription as to which boundary conditions should be selected in order to obtain physically-relevant results pertaining to confined higher spin fields. This lack of understanding is significant given that boundary-dependent phenomena are ubiquitous across physics, a prominent example being the Casimir effect. Here, we use the two-spinor calculus formalism to present a unified treatment of boundary conditions routinely employed in the treatment of spin-$1/2$ and spin-$1$ fields. We then use this unification to obtain a boundary condition that can be applied to massless fields of any spin, including the spin-$2$ graviton, and its supersymmetric partner the spin-$3/2$ gravitino. 

\end{abstract}


\maketitle

The coupling of a quantised field to matter causes the spectrum of its vacuum fluctuations to change. The range of resulting phenomena includes what are variously known as Casimir forces, energies and pressures. The simple case of two perfectly reflecting, infinite, parallel plates, that impose boundary conditions (BCs) on the Maxwell field was investigated by Casimir in \cite{casimir1948attraction}. Casimir's seminal paper has since resulted in a wide range of extensions, generalizations and experimental confirmations over the last half-century or so \cite{LandauLifshitzPitaevskii, Boyer:1968ga, Schwinger:1978be,Lamoreaux:1997cu, Mohideen:1998ip, Rahi:2009fx}. This has led, for example, to new constraints on hypothetical Yukawa corrections to Newtonian gravity \cite{DeccaYukawa}. Casimir's relatively simple and intuitive calculation has provided an enormously fruitful link between real-world experiments and the abstract discipline of quantum field theory. In fact, boundary-dependent effects are often cited in standard quantum field theory textbooks as the primary justification for the reality of vacuum fluctuations. Such interpretations however, are not without controversy \cite{Jaffe:2005km}. Boundary-dependent vacuum forces are not specific to electromagnetism, and are in fact a general feature of quantised fields. Here we provide a unified and physically well-motivated treatment of the effects that perfectly reflecting material boundaries have on \emph{any} quantum field. 

A striking example of non-electromagnetic Casimir effects can be found in nuclear physics, wherein early attempts to model the nucleon without considering BCs at its surface ran into a variety of problems \cite{ThomasWeiseStructNucleon}. Many of these problems were solved by the introduction of the `bag model' \cite{Chodos:1974de}, which describes a nucleon as a collection of free massless quarks  \footnote{This is justified because the energy scale associated with the nucleon radius is much larger than that associated with the mass of the quark} confined to a region of space (the `bag'), with a postulated BC that governs their behaviour at the surface. This model, subject to sensible choices of a small number of free parameters, correctly predicts much of the physics of the nucleon \cite{ThomasWeiseStructNucleon}. The boundary-dependent vacuum contribution to the energy (the Casimir energy) has important consequences for the stability of the bag \cite{Milton:1983ht, Milton:1983gg, Oxman:2005du}. This further emphasises the importance of using physically-motivated BCs. Another example of the need to impose physical BCs on fermionic fields is provided by graphene and carbon nanotubes, both of which are the subject of intense contemporary interest. These structures support a two-dimensional gas of massless fermions \cite{Novoselov:2005es} and the resultant fermionic Casimir force has been found to have even a different \emph{sign} depending on the precise choice of BCs, namely periodic or anti-periodic \cite{Bellucci:2009fv}. Single carbon nanotubes have been proposed as nanomechanical switches \cite{Sapmaz:2003gt} whose failure modes may include stiction caused by Casimir forces \cite{Andrews:2014cp}. 
  
 \begin{figure}[h]
\center
\includegraphics[width=0.6\columnwidth]{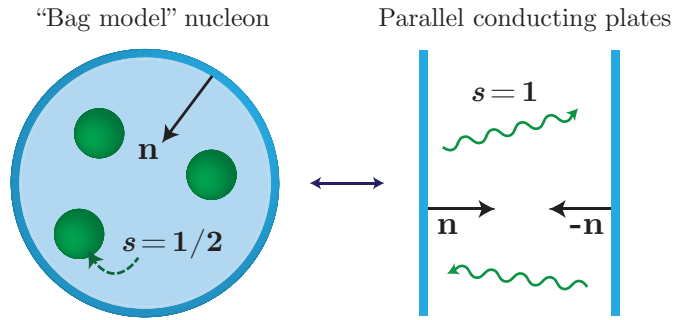} 
\caption{Schematic representation of the main idea of our work. We exploit a correspondence between the bag model of the nucleon and the electromagnetic Casimir force between parallel conducting plates. This enables us to unify them and subsequently generalise them in order to treat arbitrary spinor fields.} \label{SetupFig}
\end{figure}

Given that Casimir effects associated with the Maxwell (spin-1) field and the Dirac (spin-$\nicefrac{1}{2}$)  field are of experimental and theoretical interest, one is naturally led to the question as to whether the Casimir effect for these fields can be calculated in a unified way. Since Casimir physics is essentially the study of BCs, can we construct BCs that include those used for the spin-$\nicefrac{1}{2}$ and spin-$1$ fields as special cases? Furthermore, can we generalise this unified BC to one that applies to higher-spin fields? Answering these questions would significantly advance our understanding of the physics of confined higher-spin fields. For example, in \cite{WenBiao:2008dk} arbitrary BCs (periodic) are applied to the spin-$\nicefrac{3}{2}$ field --- no physical justification is attempted. Here we unify the BCs usually employed in the treatment of spin-$\nicefrac{1}{2}$ and spin-$1$ fields near perfect reflectors, and then develop this unification in order to model the confinement of fields possessing arbitrary spin.

We will begin our treatment by outlining the BCs assumed within the bag model, i.e., those usually employed in the treatment of massless spin-$\nicefrac{1}{2}$ particles. In this model, one envisages a fermionic field confined to some region of space that is surrounded by an impenetrable barrier. Thus, a physically reasonable constraint to impose (which can also be motivated by an appropriate choice of Lagrangian \cite{ThomasWeiseStructNucleon,Johnson:1978ip}) is that there be no particle current across the surface;
\begin{equation}
(n_\mu j^\mu) |_\mathcal{B}=0 \qquad j^\mu = \phi \gamma^\mu \psi \, , \label{DiracBC}
\end{equation}
where $n_\mu$ is a spacelike unit four-vector normal to the surface defining the bag, and where we have employed the summation convention for repeated upper and lower indices. Rather than using the usual notation $\bar{\psi}$ to denote the Dirac adjoint $\psi^\dagger \gamma^0$ of $\psi$, we have used $\phi\equiv \psi^\dagger \gamma^0$ in order to avoid confusion later on. The constraint \eqref{DiracBC} is obeyed if $\psi(x)\equiv\psi$ satisfies
\begin{equation}
i n_\mu \gamma^\mu \psi = \psi \qquad \B{x} \in \mathcal{B}  \label{DiracBCWaveEqn}.
\end{equation}
This can be shown by multiplying Eq.~\eqref{DiracBCWaveEqn} by $\phi$ from the left, and the Dirac adjoint of Eq.~\eqref{DiracBCWaveEqn}  by $\psi$ from the right. Adding these two quantities, one finds $ 2in_\mu \phi \gamma^\mu \psi=2in_\mu j^\mu=0$. This shows that the BC \eqref{DiracBCWaveEqn} implies $n_\mu j^\mu =0$, which is the constraint imposed in the bag model. 

What about higher spins? It is well-known that the description of fields with arbitrary spin can be constructed using elementary two-spinors via the so-called two-spinor calculus formalism  \cite{vanderWaerden:1928vw, Bade:1953cp, PenroseRindler, schwinger1965angular}. This means that, for example, the Maxwell field can be described on the same footing as the massless Dirac field. As a result, we should be able to find a spin-$1$ analog of the constraint \eqref{DiracBC}. Initially this might seem hopeless, because no local particle-current exists for fields with spin greater than $\nicefrac{1}{2}$ \cite{Penrose:1965hj}. However, we shall see that there is a natural adaptation of the Dirac-field BC to the Maxwell field, which moreover coincides with the BC usually employed in the calculation of the electromagnetic Casimir force. This allows us to generalise the BC \eqref{DiracBCWaveEqn} to fields with arbitrary spin.

The two-spinor calculus allows one to build irreducible representations of the universal covering group of the homogeneous Lorentz group using two-dimensional complex symplectic vector spaces $S$ and ${\bar S}$, where a bar is used to denote the complex conjugate space. The space $S$ is the pair $(V,\omega)$, where $V$ is a two-dimensional complex vector space and $\omega$ is a complex symplectic (non-degenerate) form. Choosing a basis $\{f_a\}\subset V$ we can write arbitrary elements (spinors) of $S$ and $\bar{S}$ as 
\begin{equation}
\psi = \psi^a f_a \in S \qquad {\bar \psi}=\psi^{\bar a}f_{\bar a}\in {\bar S}, 
\end{equation}
where we use bars rather than the more commonly used dots to distinguish between a spinor index and a conjugate-spinor index. Furthermore we rely entirely on the different indices in order to distinguish between the components of $\psi$ and ${\bar \psi}$. With these index conventions matrix operations become particularly simple. If a matrix $v$ has elements $v^{ab}$ then we have the following representations
\begin{equation}
v\leftrightarrow v^{ab},\quad {\bar v} \leftrightarrow v^{{\bar a}{\bar b}},\quad{v}^\intercal \leftrightarrow v^{ba}, \quad{v}^\dagger \leftrightarrow v^{{\bar b}{\bar a}}
\end{equation}
where $^\intercal$ and $^\dagger$ denote matrix transposition and Hermitian conjugation respectively. The symplectic form $\omega$ is used to raise and lower spinor indices. We adopt the convention that $\omega_{ab}=-\omega_{ba}$ can only be used to lower an index when the repeated index is in the first slot. Similarly $\omega^{ab}$ only raises the index when the repeated index is in the second slot. The same rules apply for barred indices, so altogether
\begin{equation}
\omega_{ab}\psi^a=\psi_b, \omega^{ab}\psi_b=\psi^a,  \\
\omega_{{\bar a}{\bar b}}\psi^{\bar a}=\psi_{\bar b},  \omega^{{\bar a}{\bar b}}\psi_{\bar b} = \psi^{\bar a}. \label{UppyDownyRules}
\end{equation}
We note that these identities imply the following identity for the contraction of a rank-$n$ spin tensor with its dual
\begin{equation}
\phi_{a_1 a_2...a_n} \phi^{a_1 a_2...a_n} = (-1)^n \phi_{a_1 a_2...a_n} \phi^{a_1 a_2...a_n} \label{AbsIdentity}. 
\end{equation}
This means that for odd $n$ (fermionic fields) the quantity $\phi_{a_1 a_2...a_n} \phi^{a_1 a_2...a_n}$ is identically zero. 

The above ingredients allow one to write a spacetime tensor of rank $(i,j)$ in terms of Hermitian matrices as
\begin{equation}
T^{\mu_1 ... \mu_i}_{\hphantom{\mu_1 ... \mu_i} \nu_1...\nu_j}=  
 \sigma^{\mu_1}_{\hphantom{\mu_1} \bar{a}_1 a_1}...\sigma^{\mu_i}_{\hphantom{\mu_1} \bar{a}_i a_i} \tilde{\sigma}_{\nu_1}^{\hphantom{\nu_1}{b}_1 \bar{b}_1}...\tilde{\sigma}_{\nu_i}^{\hphantom{\mu_1}{b}_i \bar{b}_i}T^{\bar{a}_1 a_1...\bar{a}_i a_i}_{\qquad \qquad \bar{b}_1 b_1...\bar{b}_i b_i} \label{TDef}
\end{equation}
where 
\begin{equation}
\sigma^\mu = (\mathbb{I}, \sigma^i),\qquad  \tilde{\sigma}^\mu = (\mathbb{I}, -\sigma^i)
\end{equation}
with $\sigma^i$ the $i$th Pauli matrix. We have now laid out a formalism that we can use to describe fields of arbitrary spin. This will eventually enable us to determine a unified physical BC applicable to any massless spinor field. We begin this process by rewriting the right-helicity component of the Dirac current in \eqref{DiracBC} as
\begin{equation}
j^\mu  = \sigma^\mu_{\;\; \bar{a} a}   j^{a\bar{a}}, \qquad   j^{a\bar{a}} \equiv \psi^{\bar{a}} \psi^a. \label{CurrentSpinorNotation}
\end{equation}
In terms of the two-spinor calculus formalism, the BC \eqref{DiracBCWaveEqn} for the right-helicity component becomes
\begin{equation}
n_\mu \sigma^\mu_{\, \, \bar{a} a} \psi^a = \psi_{\bar{a}} \qquad \B{x} \in \mathcal{B} . \label{DiracBCSpinorNotation}
\end{equation}
We can demonstrate that Eq.~\eqref{DiracBCSpinorNotation} implies $n_\mu j^\mu=0$ by multiplying both sides by $\psi^{\bar{a}}$ and using the identity \eqref{AbsIdentity}, which gives
\begin{equation}
n_\mu \sigma^\mu_{\, \, \bar{a} a} \psi^{\bar{a}}\psi^a = \psi_{\bar{a}}\psi^{\bar{a}}\equiv 0 \qquad \B{x} \in \mathcal{B}.\label{PlusCurrentSpinorNotation}
\end{equation}
This shows that Eq.~\eqref{DiracBCSpinorNotation} is indeed the two-spinor calculus version of the bag BC \eqref{DiracBCWaveEqn} for a right-helicity spinor.  A similar calculation holds for the left-helicity spinor.

 The next-lowest spin field after the Dirac field ($s = \nicefrac{1}{2}$) is of course the Maxwell field ($s=1$). Just as in our discussion of the Dirac field, we will begin by casting the usual statements of the BCs (in this case given by restrictions on the electric and magnetic fields $\B{E}$ and $\B{B}$) in the language of two-spinor calculus.  The electromagnetic BC for a perfect conductor requires that $\B{n}\times \B{E}$ and $\B{n}\cdot \B{B}$ vanish at the surface. This in turn implies that $\B{n}\cdot \B{S}$ also vanishes, where $\B{S}=\B{E}\times\B{B}$ is the Poynting vector. Using the Riemann-Silberstein (RS) vector $\B{F}\equiv\B{E}+i \B{B}$, the electromagnetic BCs can be written 
\begin{equation}
\text{Re} \left[\B{n} \times \B{F} \right]= 0, \qquad\text{Im} \left[ \B{n} \cdot  \B{F} \right] = 0 \qquad \B{x} \in \mathcal{B}. \label{BCsOnF} 
\end{equation}
We can assume without loss of generality that $n_\mu = (0,\hat{\B{z}})$ so that the RS vector obeying the BCs \eqref{BCsOnF} is
\begin{equation}
\B{F}= (i B_x, iB_y, E_z).\label{RSOnBoundary}
\end{equation}
Following \cite{BialynickiBirula:1996cf}, we now introduce the spin tensor $ \phi^{ab}$ such that 
\begin{eqnarray}
 \phi^{00} =  -F_x+iF_y \qquad   &  \phi_{\bar{0}\bar{0}} =   \bar{F}_x+i\bar{F}_y,    \\
 \phi^{01} =  F_z = \phi^{10},   \qquad & \phi_{\bar{0}\bar{1}} = \bar{F}_z =   \phi_{\bar{1}\bar{0}},  \\
  \phi^{11} =  F_x+iF_y,  \qquad & \phi_{\bar{1}\bar{1}}=  -\bar{F}_x-i\bar{F}_y,
 \end{eqnarray}
in terms of which \eqref{RSOnBoundary} can be written as
\begin{equation}
 \phi^{00} =  \phi_{\bar{0}\bar{0}},\qquad    \phi^{01} =  -\phi_{\bar{0}\bar{1}}. \label{PhiConditions}
\end{equation}
Using Eq.~\eqref{TDef}  we can write a symmetric tensor $T^{\mu\nu}$ as
\begin{equation}
T^{\mu\nu} = \sigma^\mu_{\, \, \bar{a} a} \sigma^\nu_{\, \, \bar{b} b} T^{\bar{a} a \bar{b} b} \label{TGeneral}
\end{equation}
where $T^{\bar{a} a \bar{b} b}$ is a symmetric spin-tensor. If we define  $T^{\bar{a} a \bar{b} b} = \phi^{\bar{a}\bar{b}} \phi^{ab}$, then $T^{\mu\nu}$ in Eq.~\eqref{TGeneral} is the familiar electromagnetic energy-momentum tensor, with components
\begin{equation}
T^{00} = \B{E}^2+\B{B}^2, \qquad  T^{i0} = 2 (\B{E}\times \B{B})^i = 2 S^i. 
\end{equation}
In terms of $T^{\mu\nu}$ the constraint $\B{n}\cdot \B{S}$ becomes 
\begin{equation}
n_i T^{0i} = 0    \qquad \B{x} \in \mathcal{B},
\end{equation}
which for $n_\mu = (0,\B{\hat{z}})$  can be written
\begin{equation}
n_\mu T^{\mu 0 } =0  \qquad \B{x} \in \mathcal{B}.
\end{equation}
Comparing this with \eqref{DiracBC}, we see that $T^{\mu 0}$ plays the role of the Dirac current $j^\mu$ for the Maxwell field. The physical constraint, analogous to  \eqref{DiracBC}, that we impose on the Maxwell field is therefore
\begin{equation}
 n_\mu T^{\mu 0 }  = n_\mu {\sigma^\mu}_{{\bar a}a} {\sigma^0}_{\bar{b}b} \phi^{{\bar a}\bar{b}}\phi^{ab} =0 \qquad \B{x} \in \mathcal{B}\label{MaxwellConstraint},
\end{equation}
which will necessarily hold if
\begin{equation}
 n_\mu n_\nu \sigma^\mu_{\, \, \bar{a} a} \sigma^\nu_{\, \, \bar{b} b} \phi^{ab}=\phi_{\bar{a}\bar{b}} \qquad \B{x} \in \mathcal{B} .\label{MaxwellBC}
\end{equation}
We can easily demonstrate that the BC \eqref{MaxwellBC} implies the constraint \eqref{MaxwellConstraint} by again taking $n_\mu=(0,\B{\hat{z}})$, so that the BC becomes
\begin{equation}
\sigma^3_{\, \, \bar{a} a} \sigma^3_{\, \, \bar{b} b} \phi^{ab}=\phi_{\bar{a}\bar{b}} \qquad \B{x} \in \mathcal{B}. \label{zMaxwellBC}
\end{equation}
Using the explicit form of the Pauli matrices, Eq.~\eqref{zMaxwellBC} immediately yields Eqs.~\eqref{PhiConditions}, which themselves followed from having written the BCs \eqref{DiracBCSpinorNotation} and \eqref{MaxwellBC} in terms of the RS vector. 

The fact that the above procedure is exactly analogous to that for the Dirac field is remarkable and unexpected. As already mentioned, no local particle-current exists for massless fields with spin greater than $\nicefrac{1}{2}$. However, one of the few local observables associated with photons is their energy current, which is precisely the quantity that naturally appears in the spin-$1$ constraint \eqref{MaxwellConstraint}.

The generalization of the BC to arbitrary spinor fields is now clear. For spin-$\nicefrac{m}{2}$ we write our \emph{generalized bag-like boundary condition} 
\begin{equation}
n_{\mu_1} n_{\mu_2} ...  n_{\mu_m} \sigma^{\mu_1}_{\;\;\;\, \bar{a}_1 a_1} \sigma^{\mu_2}_{\;\;\;\,\bar{a}_2 a_2}... \sigma^{\mu_2}_{\;\;\;\, \bar{a}_m a_m}\phi^{a_1a_2...a_m}  =\phi_{\bar{a}_1\bar{a}_2...\bar{a}_m}  \label{GenBC},
\end{equation}
for  $\B{x} \in \mathcal{B}$. This implies $[n_\mu \mathcal{J}^\mu (m)]|_\mathcal{B} =0 $ where
\begin{equation}
\mathcal{J}^\mu(m) = {\sigma^\mu}_{{\bar a}_1a_1} {\sigma^0}_{{\bar a}_2a_2}... {\sigma^0}_{{\bar a}_ma_m}\phi^{a_1a_2...a_m}\phi^{{\bar a}_1{\bar a}_2...{\bar a}_m}
\end{equation}
is the local current for the spinor field concerned --- the Dirac field has $\mathcal{J}^\mu(1)=j^\mu$, the Maxwell field has $\mathcal{J}^\mu(2)=T^{\mu0}$ and so on. In terms of the spin-tensor $\phi$, the current $\mathcal{J}$ is defined by
\begin{equation}
\mathcal{J}^{a_1\bar{a}_1 a_2\bar{a}_2... a_m\bar{a}_m}\equiv\phi^{a_1a_2 ... a_m}\phi^{\bar{a}_1\bar{a}_2...\bar{a}_m}  .
\end{equation}
The BC \eqref{GenBC} ensures that
\begin{equation}
\left[n_\mu \mathcal{J}^\mu(m) \right]_\mathcal{B}=0, \label{GenConstraint}
\end{equation}
We can prove this by using the rules $\eqref{UppyDownyRules}$, which allow \eqref{GenBC} to be written as
\begin{equation}
\omega^{{\bar a}_1{\bar a}_1'}{\sigma^3}_{{\bar a}_1'a_1}...\omega^{{\bar a}_m{\bar a}_m'}{\sigma^3}_{{\bar a}_m'a_m}\psi^{a_1...a_m} = \psi^{{\bar a}_1...{\bar a}_m}.
\end{equation}
Substituting this into $n_\mu \mathcal{J}^\mu$ and using the explicit forms of the Pauli matrices along with the matrix representation $\omega=i\sigma^2$, we find
\begin{eqnarray}
n_\mu \mathcal{J}^\mu(m)  &= (-1)^m {\sigma^1}_{a_2 a_2'}...{\sigma^1}_{a_m a_m'}\psi^{a_1'...a_m'}{\psi_{a_1'}}^{a_2...a_m} \nonumber \\
&= -(-1)^m  {\sigma^1}_{a_2 a_2'}...{\sigma^1}_{a_m a_m'}{\psi_{a_1'}}^{a_2'...a_m'}\psi^{a_1'a_2...a_m}. \label{ProofLastLine}
\end{eqnarray}
Using $\sigma^1 = (\sigma^1)^\intercal$ and relabelling the indices $a_i \leftrightarrow a_i'$ for $i=2,...,m$,  the last line of Eq.~\eqref{ProofLastLine} is equal to
\begin{equation}
n_\mu \mathcal{J}^\mu(m) =(-1)^m  {\sigma^1}_{a_2 a_2'}...{\sigma^1}_{a_m a_m'}\psi^{a_1'...a_m'}{\psi_{a_1'}}^{a_2...a_m},
\end{equation}
which is the negative of the first line in Eq. \eqref{ProofLastLine}. This proves that the BC \eqref{GenBC} implies $n_\mu \mathcal{J}^\mu(m) =0$ for an arbitrary spin-$\nicefrac{m}{2}$ field, meaning that it is indeed a generalized bag-like boundary condition and is the main result of our work.

As we have already noted, the identification of a physical current $\mathcal{J}$ for higher spin fields seems at first problematic, due to the non-existence of a local particle current for spin $>\nicefrac{1}{2}$. We have in fact already tackled this problem by adapting the spin-$\nicefrac{1}{2}$ BCs to the spin-$1$ case. This enables us to inductively determine the appropriate $\mathcal{J}$ for higher spin fields.

 Particularly noteworthy is identification of $\mathcal{J}$ for the spin-$2$ field that describes linearised quantum gravity. This field is most commonly described using a symmetric traceless tensor field $h^{\mu\nu}$ that results from the first-order expansion $g^{\mu\nu}(u)=g^{\mu\nu}+uh^{\mu\nu}+...\,$ of the general metric tensor of curved spacetime. In this first-order approximation, Einstein's vacuum equations in terms of $h^{\mu\nu}$ are equivalent to the correct relativistic wave equation for a massless spin-$2$ particle (the so-called graviton). The right and left-helicities of the graviton are described by symmetric spin-tensors $\psi^{abcd}$ and $\psi_{{\bar a}{\bar b}{\bar c}{\bar d}}$ respectively. These can be used to define the Bel-Robinson tensor, which is a strong candidate for the gravitational version of a symmetric energy-momentum tensor \cite{PenroseRindler}. While it is well-known that the gravitational field does not possess a unique local energy-momentum tensor, the Bel-Robinson tensor $T^{\mu\nu\rho\sigma}$ possesses many of the properties usually associated with such objects, namely, total-symmetry, tracelessness and certain positivity properties \cite{PenroseRindler}. It is also the natural spin-$2$ analog of the symmetric energy-momentum tensor $T^{\mu\nu}$ of electrodynamics. The generalised BC in Eq. \eqref{GenBC} therefore implies the vanishing of the local current $T^{\mu 000}$. Analogously to the currents encountered in the spin-$\nicefrac{1}{2}$ and spin-$1$ cases, $T^{\mu 000}$ could be viewed as a natural quantity in terms of which the physical BC should be specified for the spin-$2$ field. 

A possible impact of our unified BC is the ability to transfer well-known techniques from electromagnetism to fields with different spin. This could prove especially fruitful in extending our work to consideration of imperfectly reflecting boundaries, as was done very recently in \cite{Quach:2015ev} for the particular case of the spin-$2$ graviton. 

To conclude, we have reported the first unified treatment of physical (bag-like) boundary conditions at perfect reflectors for fields with arbitrary spin. This was achieved by writing well-known BCs for the Maxwell and massless Dirac fields in a unified language, and then carrying out a natural generalisation. The very existence of a unified BC for the Maxwell and Dirac fields is a remarkable result on its own because of the fundamental differences between the conserved currents for the two fields. However, we have shown that such a BC does exist --- the unification of two apparently disparate approaches within one self-consistent model is a satisfying result, but moreover the unification proceeds in such a way that it can be naturally extended to find completely new bag-like BCs for fields with any spin, meaning that our work opens up a whole landscape of study in confinement of higher-spin fields. 

Acknowledgements: It is a pleasure to thank Almut Beige and Jiannis Pachos for helpful comments. Financial support from the UK Engineering and Physical Sciences Research Council (EPSRC) is greatly appreciated.

\end{document}